\documentclass[conference]{IEEEtran}
\IEEEoverridecommandlockouts
\usepackage{cite}
\usepackage{amsmath,amssymb,amsfonts}
\usepackage{algorithmic}
\usepackage{graphicx}
\usepackage{textcomp}
\usepackage{xcolor}

\usepackage{subfigure}
\usepackage{subfig}
\usepackage{placeins}

\usepackage{stfloats}
\usepackage{multirow}

\def\BibTeX{{\rm B\kern-.05em{\sc i\kern-.025em b}\kern-.08em
    T\kern-.1667em\lower.7ex\hbox{E}\kern-.125emX}}
\begin{document}

\title{Semi-supervised Learning Framework for UAV Detection\\

\thanks{This work has been supported in part by NASA under the Federal Award ID number NNX17AJ94A, and by NSF CNS-1939334 Aerial Experimentation Research Platform for Advanced Wireless (AERPAW) project that supported the experiments at NC State.}
}

\author{
\IEEEauthorblockN{Olusiji O Medaiyese\IEEEauthorrefmark{1},
Martins Ezuma\IEEEauthorrefmark{2},
Adrian P Lauf\IEEEauthorrefmark{1} and
Ismail Guvenc\IEEEauthorrefmark{2} 
}

\IEEEauthorblockA{\IEEEauthorrefmark{1}Department of Computer Science and Engineering\\
University of Louisville,  Louisville, Kentucky,  40292, USA\\  }
\IEEEauthorblockA{\IEEEauthorrefmark{2}Department of Electrical and Computer Engineering\\
North Carolina State University
Raleigh, North Carolina, 27606, USA\\
Email: o0meda01@louisville.edu, mcezuma@ncsu.edu,~aplauf01@louisville.edu,~iguvenc@ncsu.edu}
}

\maketitle

\begin{abstract}
The use of supervised learning with various sensing techniques such as audio, visual imaging, thermal sensing, RADAR, and radio frequency (RF) have been widely applied in the detection of unmanned aerial vehicles (UAV) in an environment. However, little or no attention has been given to the application of unsupervised or semi-supervised algorithms for UAV detection. In this paper, we proposed a semi-supervised technique and architecture for detecting UAVs in an environment by exploiting the RF signals (i.e., fingerprints) between a UAV and its flight-controller communication under wireless inference such as Bluetooth and WiFi. By decomposing the RF signals using a two-level wavelet packet transform, we estimated the second moment statistic (i.e., variance) of the coefficients in each packet as a feature set. We developed a local outlier factor model as the UAV detection algorithm using the coefficient variances of the wavelet packets from WiFi and Bluetooth signals. When detecting the presence of RF-based UAV, we achieved an accuracy of 96.7$\%$ and 86$\%$ at a signal-to-noise ratio of 30~dB and 18~dB, respectively. The application of this approach is not limited to UAV detection as it can be extended to the detection of rogue RF devices in an environment.
\end{abstract}


\section{Introduction}
The application of unmanned aerial vehicles (UAVs) cannot be overemphasized and it has cut across environmental and disaster management \cite{coveney2017lightweight},  agriculture \cite{alsalam2017autonomous}, logistics and other fields. However, it is posing both security and privacy concerns as it is now being used for cybercrimes or to violate airspace regulations \cite{rattledrone2016,smuggler_drone2018}. For instance, more than a hundred UAVs are sighted in a restricted area across the United States every month through human efforts \cite{faa_uas_sight_reporting2020}.

Several detection techniques (i.e., audio,  visual imaging,  thermal sensing, RADAR, and radio frequency (RF)) have been used in detecting the presence of UAV in an environment. The pros and cons of these techniques are discussed in \cite{medaiyese2021wavelet}. We adopt RF-based detection techniques because: (i) the operation is stealthy;  (ii) it is not limited by the size of the UAV; and (iii) it can detect UAV at line-of-sight and non-line-of-sight. However, several other RF devices such as Bluetooth and WiFi devices operate in the same RF band (i.e., 2.4~GHz) as UAVs. This poses a limiting factor to this technique. The ubiquity of RF signals such as Bluetooth and WiFi cannot be denied because of their wide applications in different domains. Therefore, a UAV RF signal can be detected as an anomaly or rogue signal from other RF signals operating at 2.4~GHz. 

In the literature, both the transient and steady state of the RF signals has been considered for extracting unique signatures (i.e., RF fingerprints) for UAV detection and identification. The transient state of RF signal occurs during power on and off of devices \cite{hall2003detection}. Conversely, the steady state occurs between the interval of the start and end of transient state \cite{kennedy2008radio}. The transient state of RF signal has a short period \cite{hall2003detection} and its efficiency is degraded by a low signal to noise ratio (SNR) when compared to the steady state \cite{medaiyese2021wavelet}. However, the high variance attribute of the transient state makes it a common state for extracting RF fingerprints \cite{ ali2019assessment, ezuma2019detection,luo2020extraction,medaiyese2021wavelet}.

In \cite{ezuma2019detection,al2019rf,medaiyese2021wavelet, nemer2021rf, swinney2020unmanned, xu2020rf,zhao2018classification, zhou2018unmanned}, supervised learning approaches were adopted for UAV detection by using RF signals. The supervised learning algorithms such as Auxiliary Classifier Wasserstein Generative Adversarial Networks (AC-WGANs), deep neural network (DNN), support vector machines (SVM), k-nearest neighbor (kNN), and so on have been widely used. State-of-the-art results were achieved in terms of detection accuracy. However, a lot of effort is required in collecting RF signals from various UAVs leading to a high cost of labeling samples.

In this work, we proposed a novel detection model using commonplace signals (i.e., Bluetooth and WiFi) to detect the presence of UAV in an environment. To the best of our knowledge, this is the first concept where a semi-supervised algorithm is adopted for UAV  or rogue RF device detection research.

Our contributions can be summarized as follows:\looseness=-1

\begin{enumerate}

\item We use a two-level wavelet packet transform (WPT) to decompose the transient state of the RF signals (i.e., Bluetooth and WiFi signals). The statistical variance of the coefficients in each packet is computed as the feature set or the RF fingerprints, which is used to model a local outlier factor algorithm that detects UAV control signals.

\item We proposed a novel detection model based on semi-supervised learning for UAV detection. A local outlier factor (LOF) algorithm is trained using commonplace RF signals. In our case, Bluetooth and WiFi signals are used as the commonplace signals in the environment under surveillance. These commonplace signals are user's recognized signals that we consider as inliers. Conversely, the UAV control signals are considered as outliers which the LOF model can detect or classify as anomalous signals.

\item We present experimental results to show the performance of the UAV detection algorithm and the impact of signal-to-noise ratio on the performance of the model.

\end{enumerate}
The paper is organized as follows. Section~\ref{two} provides an overview of the UAV detection system model. Section~\ref{three} provides information about the experimental setup for data collection. Section~\ref{four} describes the methodology adopted in this work. In Section~\ref{five}, the experimental results are discussed. Finally, we give the conclusion and future work in Section~\ref{six}.

\section{System Model Overview} \label{two}
\begin{figure*}[t]
\center{\includegraphics[ clip,scale=0.48]{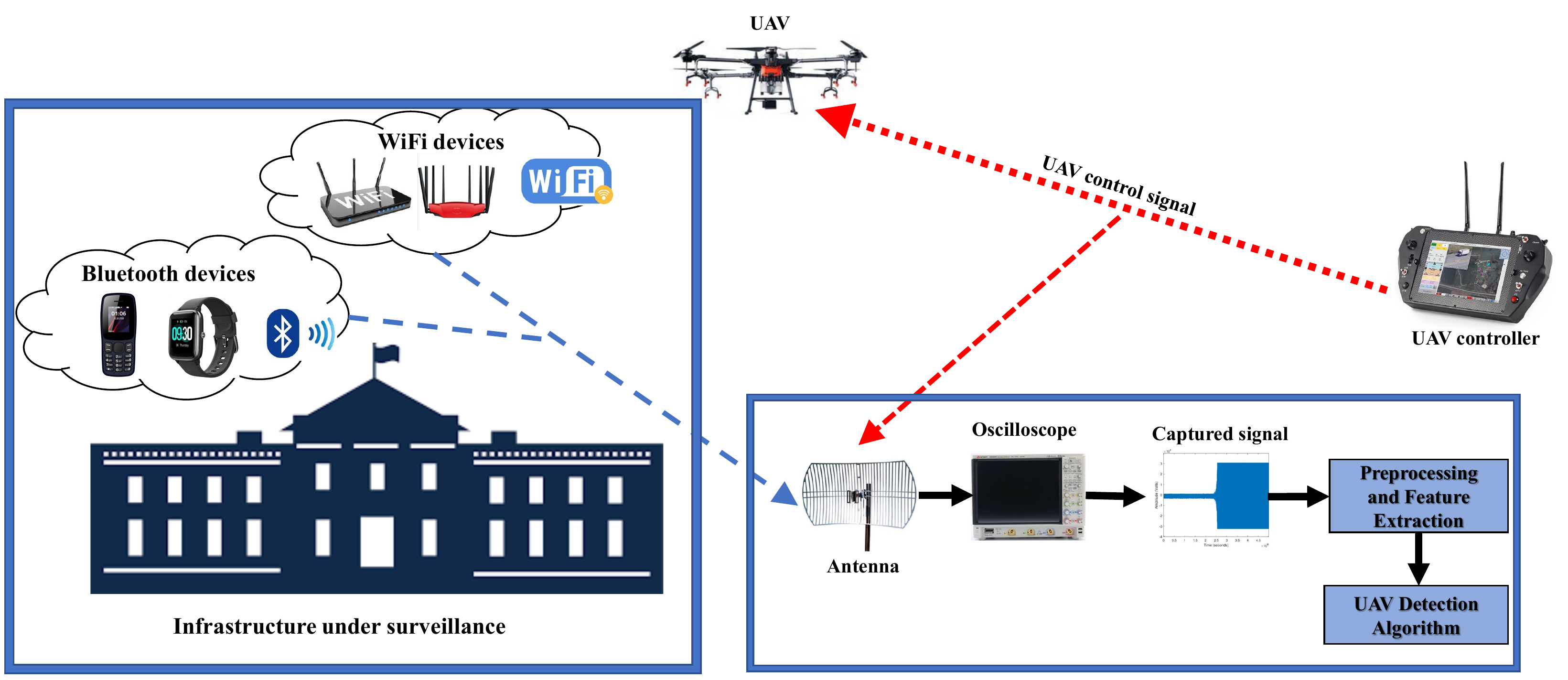}}
\caption{The scenario of the RF-based UAV detection system for infrastructure surveillance. The Bluetooth and WiFi signals in the infrastructure under surveillance are captured and preprocessed to build a UAV detection model. The detection algorithm can then classify the Bluetooth and WiFi signal as inlier signals and the UAV control signal as an outlier or anomaly signal.}
\label{Fig:system_model}
\vspace{-3mm}
\end{figure*}

Fig.~\ref{Fig:system_model} shows the system model for the RF-based UAV detection system. In the infrastructure under surveillance (IUS), both the WiFi and Bluetooth devices operate at the same frequency band (i.e., 2.4~GHz) as UAV flight-controller. The WiFi and Bluetooth signals in the environment are captured and labeled as recognized (inlier or normal) signals which are allowed to propagate around or within the IUS. The recognized signals are preprocessed to capture the transient state of the signals for feature extraction. The extracted feature set is then used to train a LOF algorithm for the detection of recognized signals (i.e., WiFi and Bluetooth) and UAV control signals.

\section{Experimental Setup and Data Capturing}\label{three}

\subsection{Data capturing step}





\begin{table}
\centering
\caption{ Catalog of RF devices used in the experiment for RF fingerprints acquisition. Under the UAV device, we use the UAV controllers from respective models.}
\label{UAV_catalogue}
\begin{tabular}{|c|c|c|}
\hline
Device & Make & Model\\
\hline
UAV &\multirow{4}{*}{\text{ DJI}} & Phantom 4  \\
&&Inspire  \\
&&Matrice 600  \\
&&Mavic Pro 1  \\\cline{2-3}

& Beebeerun & FPV RC drone mini quadcopter\\\cline{2-3}

& 3DR & Iris FS-TH9x\\
\hline
\multirow{2}{*}{\text{Bluetooth }}& Apple & Iphone 6S\\\cline{2-3}
& FitBit  &  Charge3 smartwatch\\
\hline
\multirow{2}{*}{\text{WiFI}} &Cisco & Linksys E3200\\\cline{2-3}
&TP-link & TL-WR940N\\

\hline
\end{tabular}
\end{table}

\begin{figure}[b!]
\center{\includegraphics[trim=0.1cm 0cm 0.1cm 0.1cm, clip,scale=0.45]{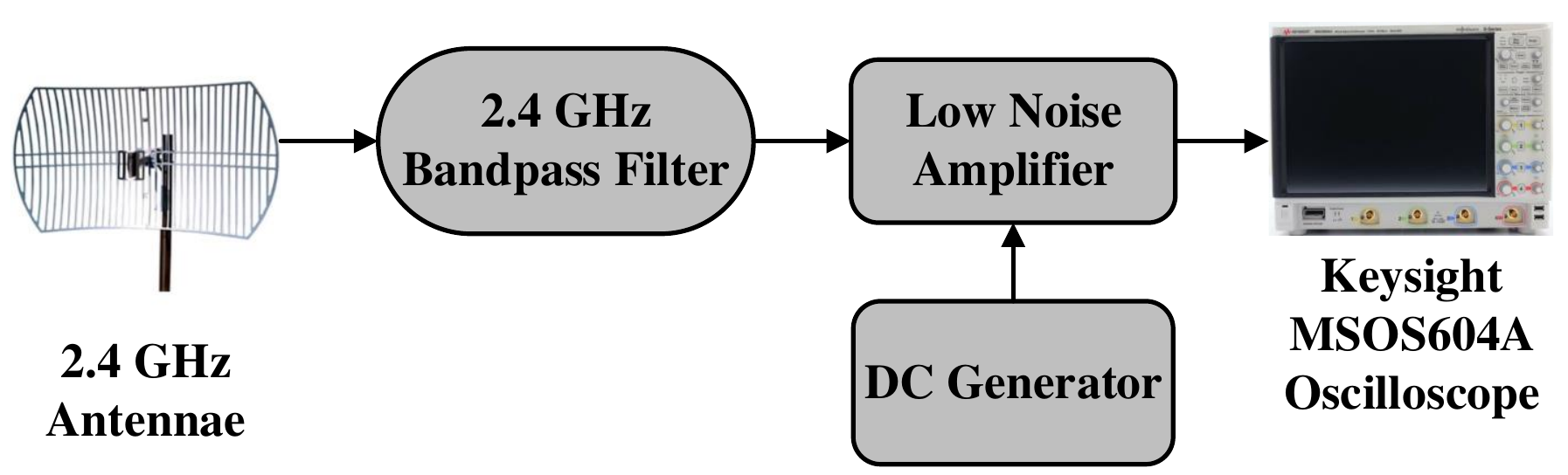}}
\caption{Data capturing schematic diagram used for experimental setup.}
\label{Fig:experimental_setup}
 \vspace{-5mm}
\end{figure}
We conducted the experiment in an outdoor setting. The devices (i.e., UAV controllers, Bluetooth, and WiFi devices) used in this experiment operate in the 2.4 GHz band. The list of RF devices used for the experiment is in Table~\ref{UAV_catalogue}. Six UAV controllers, two WiFi routers, and two Bluetooth devices are used in this experiment for data acquisition. 

Fig.~\ref{Fig:experimental_setup} illustrates the experimental setup for collecting each device's RF signals. A 2.4 GHz antenna is used for the reception of signals emitting or propagating from the devices. A 2.4 GHz bandpass filter is used to filter out signals that are not in the 2.4 GHz frequency band. The bandpass signal is amplified with a low noise amplifier. A direct current (DC) generator powers the amplifier. The amplified signal is sampled, captured, and stored using a 6 GHz bandwidth Keysight MSOS604A oscilloscope which has a sampling frequency of 20 GSa/s. 

\begin{figure}
\center{\includegraphics[trim=0.1cm 0cm 0.1cm 0.1cm, clip,scale=0.3]{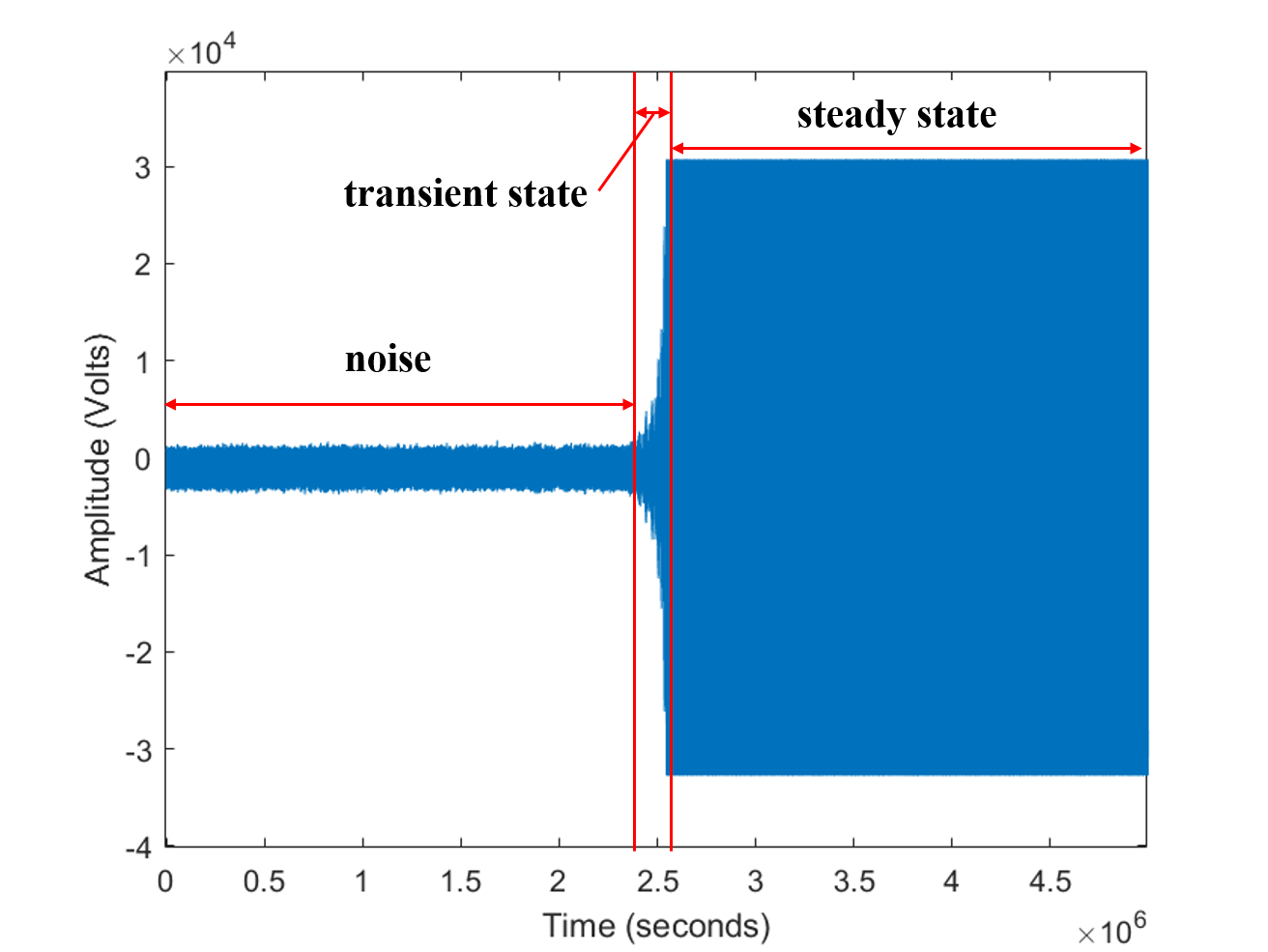}}
\caption{RF Signal from DJI Phantom controller.}
\label{Fig:signal_label}
 \vspace{-5mm}
\end{figure}

The oscilloscope has an energy threshold to differentiate the presence of any signal from background noise. When the energy level of the environment is below the threshold, the oscilloscope will be idle. This implies that there is an absence of a signal in the environment. However, when the energy level is equal to or greater than the threshold, the oscilloscope starts capturing signals. Fig.~\ref{Fig:signal_label} shows an example of signals captured from a DJI Phantom 4 controller with each part of the signal labeled. For each device in Table~\ref{UAV_catalogue}, we collected 300 RF signals at 30~dB SNR. 

\section{Methodology} \label{four}
\subsection{Feature Extraction Using Wavelet Packet Transform}

 \begin{figure}[h!]
\center{\includegraphics[ clip,scale=0.5]{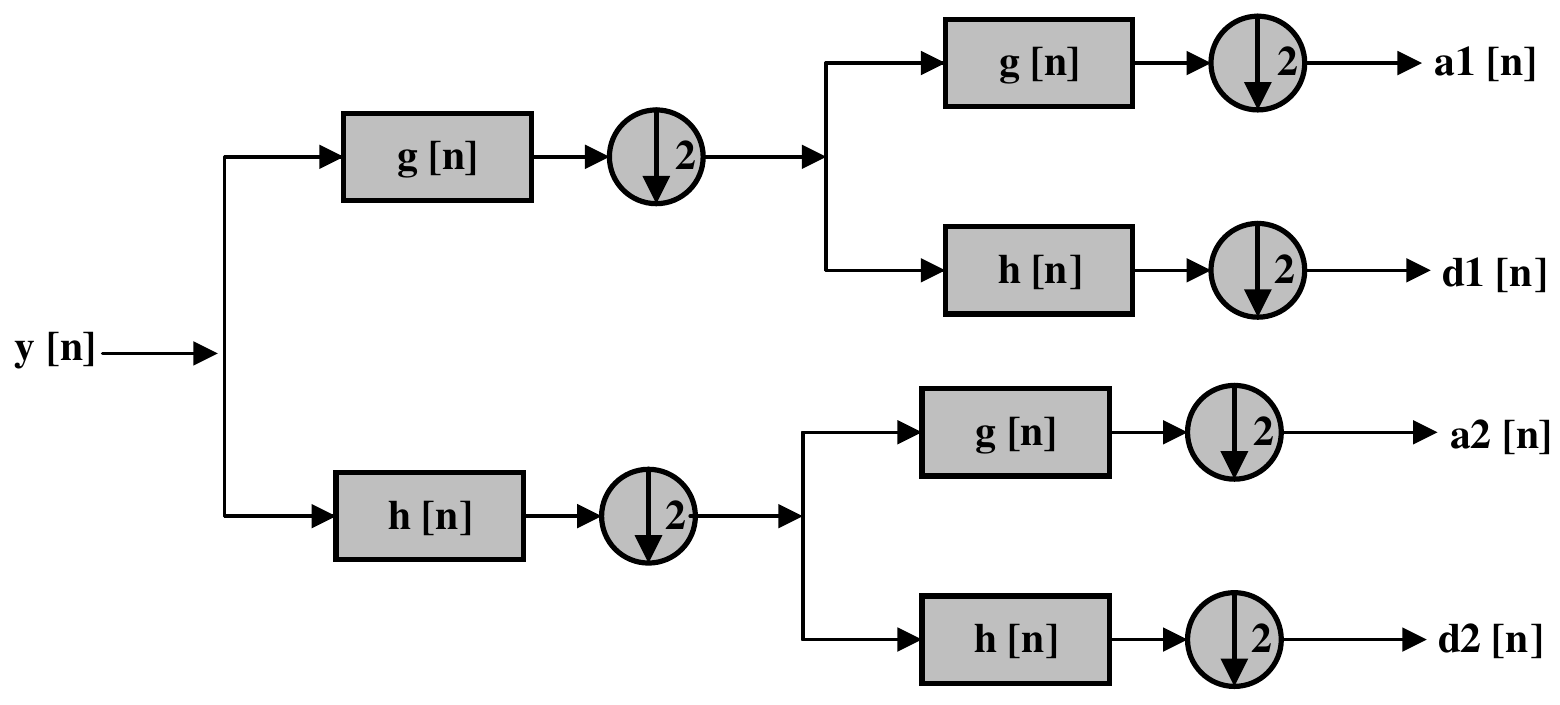}}
\caption{A two-level wavelet packet decomposition tree use to sub-band captured RF signals into packets for feature extraction. }
\label{Fig:wpt}
\vspace{-2mm}
\end{figure}

In this section, we provide a brief overview of wavelet packet transforms for feature extraction. The full details of wavelet transform are beyond the scope of this paper and can be found in \cite{ rioul1991wavelets}. The wavelet transform is the convolution of a signal $x(t)$ with a mother wavelet function $\psi_{a,b}(t)$.
The wavelet function $\psi_{a,b}(t)$ is expressed by:

\begin {equation} \label{eq:1}
\psi_{a,b}(t)= \psi( \frac{t-b}{a}).
\end{equation}

Hence, the wavelet transform of signal $x(t)$ can be mathematically defined as in (\ref{eq:2}).
\begin {equation} \label{eq:2}
W(a,b)=\frac {1}{\sqrt{a}}\int_{-\infty}^{\infty} x(t)  \psi( \frac{t-b}{a}) dt,
\end{equation}
where $a>0$ is the scaling factor which scale the wavelet function $\psi_{a,b}(t)$,  $\psi( \frac{t-b}{a})$ is the template function-base wavelet and $b$ is the time shifting factor.

Wavelet transform is generally used for multi-resolution analysis where a signal is decomposed into approximation and detail coefficient. Common variants of wavelet decomposition are the discrete wavelet transform (DWT) and the WPT. DWT sub-bands signals into approximation and detail spaces. Depending on the number of decomposition levels, only the approximation space is decomposed or split into another approximation and detail spaces. Conversely, in WPT, both the approximation and detail space are decomposed into another approximation and detail sub-space successively. The sub-spaces which are sub-band signals at a different frequency are called packets. The decomposition of both the approximation and detail spaces in WPT makes WPT have a richer time-frequency domain analysis for signals.

Fig.~\ref{Fig:wpt} shows a two-level wavelet packet decomposition. The captured signal $y[n]$ is decomposed by passing it into two parallel filters (low pass filter ($g[n]$) and high pass filter ($h[n]$)). The bandpass signal from each filter is down-sampled, decomposed further using another parallel-connected filter set and then down-sampled to give packet $a1[n]$, $d1[n]$, $a2[n]$, and $d2[n]$, respectively. Each packet preserves the global energy when the orthogonal wavelet function is adopted in decomposition. Hence, we adopt Haar wavelet as the basis function for the wavelet packet decomposition of our signal.

Eleven statistical parameters such as mean, standard deviation, mean root, absolute mean, skewness, kurtosis, variance, entropy, peak value, range, and absolute peak values are computed for each of the packets. This gives a total of 44 ($11\times4$) features extracted from the signals. To ensure a faster and cost-effective learning process, and to avoid overfitting, feature selection was done by using a ranking algorithm to pick the top four subsets from the 44 feature sets \cite{kantardzic2011data}. The metric for the ranking algorithm is based on variance. The second moment statistics of the packets are ranked. The second moment statistic, $\sigma$, gives a measure or an estimate of the coefficients spread about the mean value in a packet. It is expressed as in (\ref{eq:3}).

\begin {equation} \label{eq:3}
\sigma =\frac{\Sigma(x_j-\Bar{x})}{n-1},
\end{equation}

The statistical variance, $\sigma$, of each packet (i.e., $a1[n]$, $d1[n]$, $a2[n]$, and $d2[n]$) is used as the feature set, X,

\begin {equation} \label{eq:4}
X = (\sigma_1,\sigma_2, \sigma_3, \sigma_4),
\end{equation}
where $\sigma_1,\sigma_2, \sigma_3, \sigma_4$ are the variance of $a1[n]$, $d1[n]$, $a2[n]$, and $d2[n]$, respectively.

\begin{table*}[t!]
\setlength{\tabcolsep}{3pt}
\centering

\caption{Average accuracy of LOF model based on the validation and test set when increasing the number of neighbors for local density estimation at 30 dB SNR.}
\label{Table_accuracy}
\begin{tabular}{|c|c|c|c|c|c|c|c|c|c|c|c|c|c|c|c|c|c|c|c|c|c|}

\hline
 Metric & \multicolumn{20}{c|}{number of neighbors}\\
\cline{2-21}  
 &  10 & 20 & 30 & 40 &  50 & 60 & 70 & 80 & 90 & 100 &110& 120 & 130 & 140 & 150 & 160 & 170 & 180 & 190 & 200\\

 \hline
Validation Accuracy  &
43.9 &
95.1 &
95 &
94.7 &
94.1 &
94.7 &
95 &
96.2 &
96.5 &
\textbf {96.7} &
96.5 &
96.5 &
96.4 &
96.5 &
96.5 &
96.4 &
96.2 &
95.8 &
95.8 &
95.8 \\ 	

\hline
Test Accuracy & 45&
95.4&
95.8&
96&
95.4&
95.3&
95.6&
96.8&
96.6&
\textbf {96.7}&
96.7&
96.2&
96.5&
96.5&
96.7&
96.4&
97&
96.8&
96.8&
96.8
	 \\
 \hline
\end{tabular}
\end{table*}
\subsection{UAV Detection Algorithm}
\begin{figure}[h!]
\center{\includegraphics[ clip,scale=0.5]{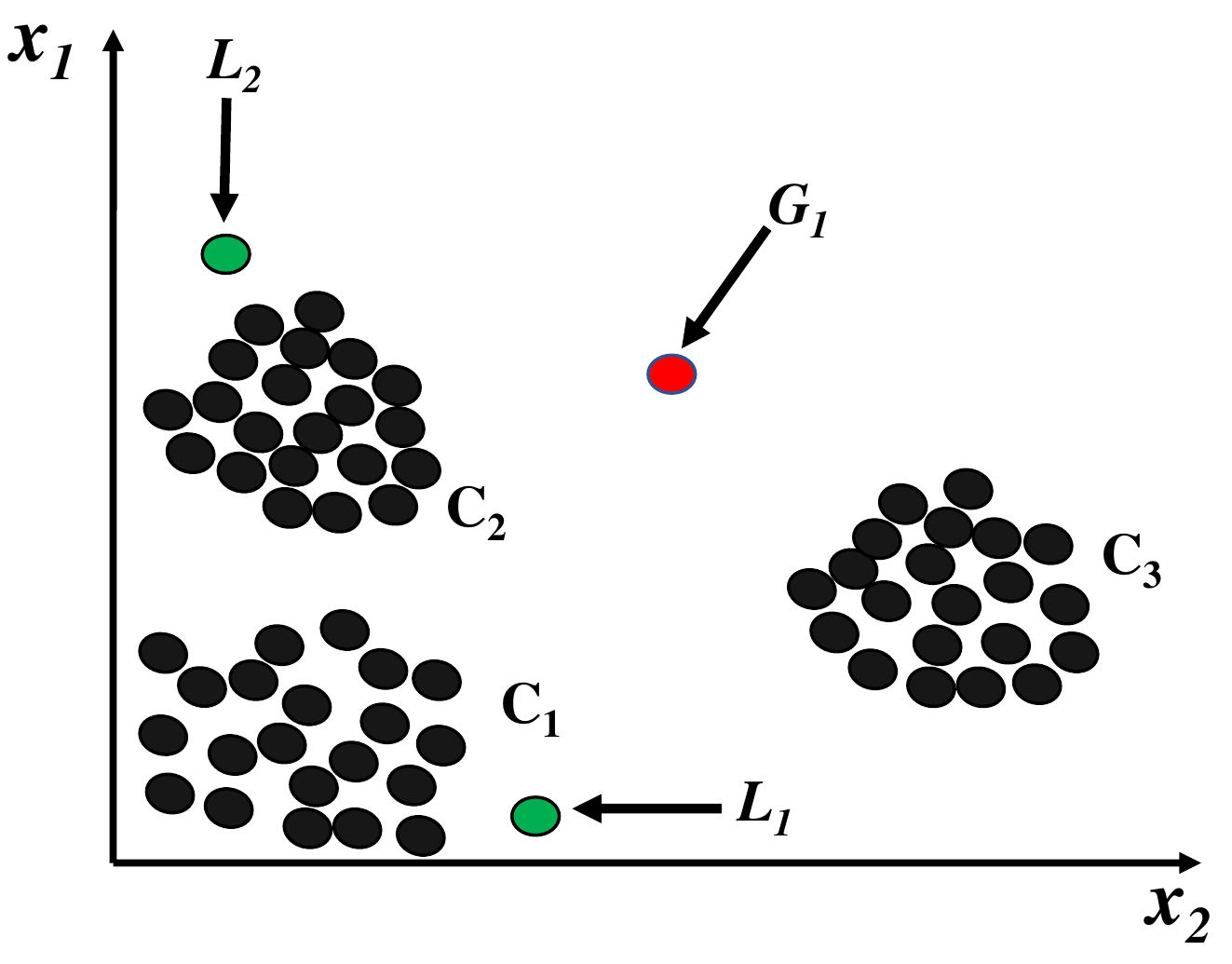}}
\caption{Illustration of local and global outliers in a dataset with three clusters. }
\label{Fig:lof}
\vspace{-2mm}
\end{figure}

Semi-supervised anomaly detection is an approach where the training set contains only the normal data (i.e., inlier) and the test set (i.e., unlabeled data ) is tested against the reference normal data in the training set by assigning anomaly score \cite{chandola2010anomaly}. The proposed model is trained using only the commonplace signals in the environment under surveillance. We adopt the LOF algorithm as the UAV detection algorithm. LOF is an algorithmic approach of assigning each datapoint in a dataset a degree of being an anomaly or outlier by measuring the local deviation of the datapoint from its neighbors \cite{breunig2000lof}. The key advantage of LOF from other outlier detection algorithms is that instead of considering outliers globally in a dataset, it takes account of an outlier from a local perspective.  Fig. \ref{Fig:lof} shows an illustration of  local ($L_1, L_2$) outliers and a global outlier ($G_1$) in a dataset that has three clusters (i.e., $C_1, C_2$ and $C_3$). $L_1$ and $L_2$ are local outliers to cluster $C_1, $ and $C_2$, respectively. Conversely, $G_1$ is a global outlier to the three clusters. A global outlier detection algorithm might not be able to detect $L_1$ and $L_2$ as outliers. Information about how LOF computes the local density of a datapoint can be found in \cite{breunig2000lof}. The implementation of the LOF algorithm in scikit-learn \cite{scikit-learn} is used for this work.

For the Bluetooth and WiFi devices, we selected 200 RF signals from each device as the training set and 100 RF signals as the evaluation set (i.e., test and validation). A total of 800 RF signals (non-UAV signals) is used to train the LOF algorithm. It is important to note that UAV control signals are not needed for training purposes (i.e., UAV control signals are not part of the train set).
The evaluation set consists of 400 non-UAV signals (i.e., Bluetooth and WiFi signal) and 1800 UAV control signals. The evaluation set is randomly split into the ratio of $70\%$ to $30\%$ for the test and validation set, respectively. To ensure the model is properly trained, we use the validation set to fine-tune the model and test the model using the test set.



\section{Experimental Results and Discussions}\label{five}

We present the results of the model during and after the training process of the algorithms. We focus our results on three main points. These points are: 
\begin{itemize}
\item selection of the number of neighbors to be used for LOF to classify signals;
\item performance metric of the detection algorithm;
\item model performance under varying SNR.
\end{itemize}

Detecting an inlier (non-UAV) or outlier (UAV) signal by the LOF model can be viewed as a two-class problem. Accuracy, precision, recall, and F1-score are examples of metrics for evaluating the performance of a classifier. The definition of these metrics is given below:

\begin {equation} \label{eq:7}
\text{Accuracy}=\frac{ T_{\rm{P}}+T_{\rm{N}}} {T_{\rm{P}}+T_{\rm{N}}+F_{\rm{P}}+F_{\rm{N}}},
\end{equation}

\begin {equation} \label{eq:8}
\text{Precision}= \frac{T_{\rm{P}}}{T_{\rm{P}}+F_{\rm{P}}},
\end{equation}

\begin {equation} \label{eq:9}
\text{Recall}= \frac{T_P}{T_P+F_N} ,
\end{equation}

\begin {equation} \label{eq:10}
{F_{\rm{1}}}~\text{score}=2\left( \frac{Precision\times Recall}{Precision + Recall}\right),
\end{equation}
where $T_{\rm {P}}$, $T_{\rm {N}}$, $F_{\rm {P}}$ and $F_{\rm {N}}$ represent true positive, true negative, false positive and false negative, respectively.

The Manhattan distance measure is used for the distance computation by the LOF model. An important factor in LOF is the number of neighbors, $n$, for estimating the local neighborhood (i.e., local density) of the datapoint or variable. The validation data is used to select or estimate the appropriate number of neighbors, $n$. Using average accuracy as a metric, Table~\ref{Table_accuracy} shows the performance of the LOF model when varying the number of neighbors, $n$, at an incremental step of 10. At $n$ equal to 100, the highest accuracy was obtained for the validation set which stands at $96.7\%$. For this reason, the number of neighbors for local density estimation used by the LOF model is 100.

\begin{figure}[h]
\center{\includegraphics[ clip,scale=0.4]{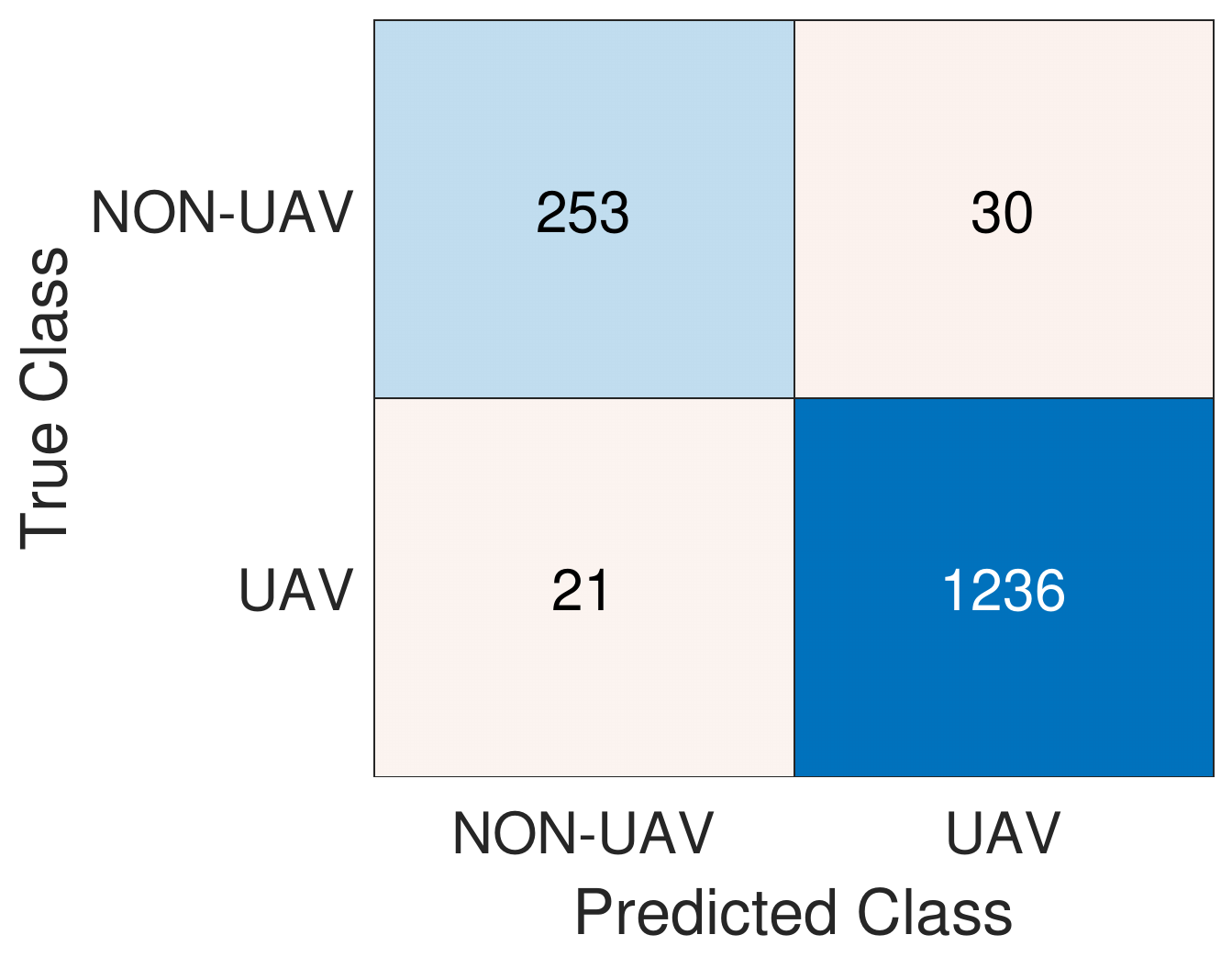}}
\caption{Confusion matrix of the LOF model ($n=100$) when evaluating the model with the test set at 30~dB. }
\label{Fig:confusion_matrix}
 \vspace{-3mm}
\end{figure}

We tested the model when $n$ is 100 for UAV detection model using the test set. The same accuracy (i.e., $96.7\%$) as the validation set is obtained for the test set. The confusion matrix of the model using the test set for evaluation is shown in Fig.~\ref{Fig:confusion_matrix} at 30 dB SNR.

\subsection{Performance at difference SNR}
\begin{figure}[h]
\center{\includegraphics[ clip,scale=0.55]{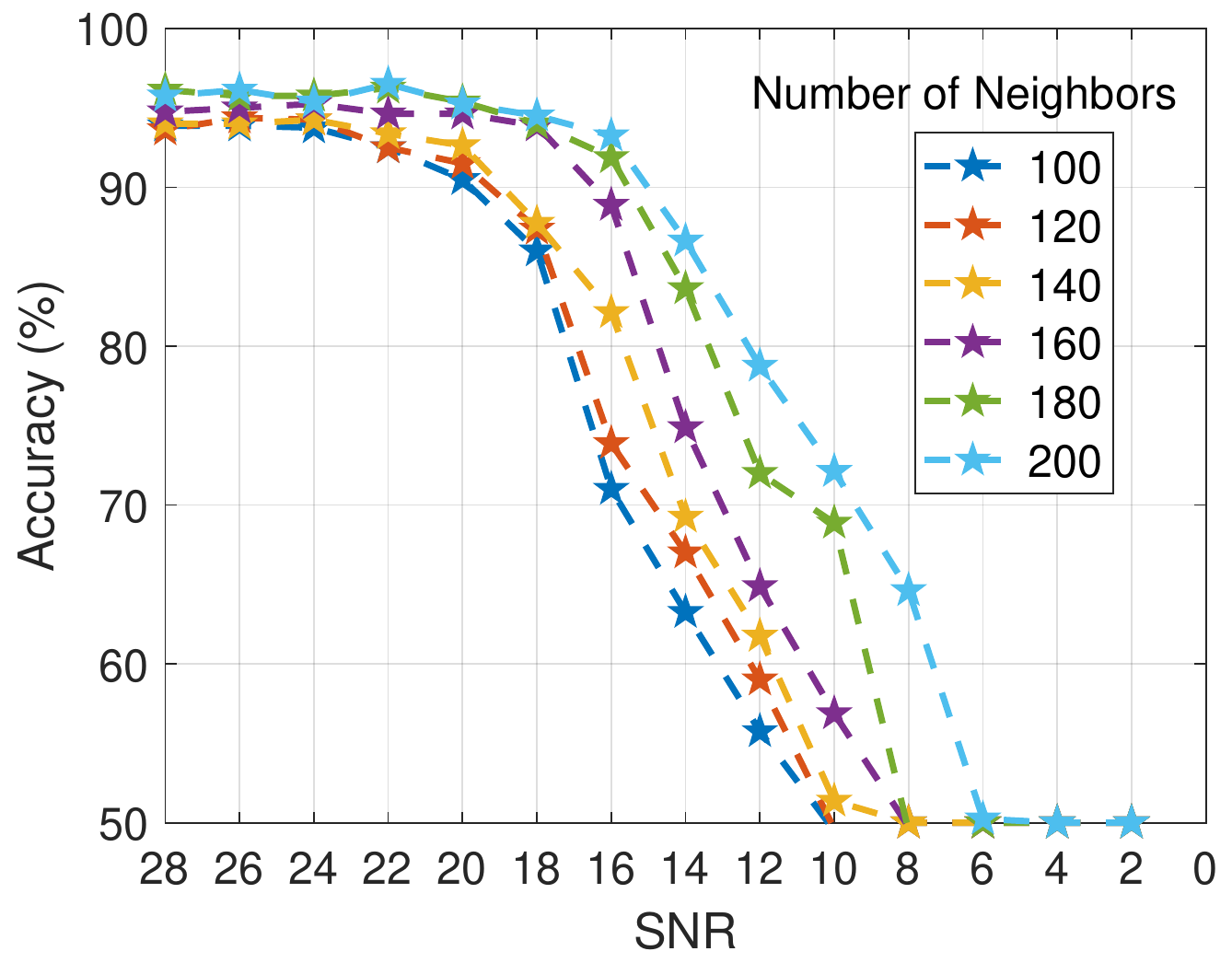}}
\caption{ Classification accuracy under varying SNR and number of neighbors for the LOF model. }
\label{Fig:snr}
\vspace{-3mm}
\end{figure}
To ensure a proper evaluation, we used a balanced test set here (i.e equal number of UAV and non-UAV signals). 200 signals from each signal type are used. Fig.~\ref{Fig:snr} shows the performance of the UAV detection algorithm at varying SNR with different number of neighbors. The classification accuracy decreases with a decrease in SNR irrespective of the number of neighbors used for estimating the local density of a datapoint. However, it was observed that the more the number of neighbors the better the model performed at a low SNR (i.e from 10~dB to 28~dB). In Fig.~\ref{Fig:snr}, from 10~dB to 28~dB SNR, the accuracy of the UAV detection algorithm increased with an increase in the number of neighbors. For instance, at 12 dB SNR, when the number of neighbors is 100, 120, 140, 160, 180, and 120, the accuracy of the model is $55.75\%$, $59\%$, $61.8\%$, $64.9\%$, $72\%$, and $78.8\%$, respectively.

More so, when the number of neighbors is given as 100, at 10~dB SNR, any signal that goes through the model is classified as a UAV control signal and the model prediction can be viewed as a random guess (i.e., accuracy is $50\%$). Conversely, at 200 neighbors, the model attained an accuracy of $72.1\%$ at 10~dB SNR and did not make a random guess until the SNR is equal to or below 6~dB.

\section{Conclusion} \label{six}
In this work, we introduced the concept of semi-supervised learning (i.e., novelty detection algorithm) for an RF-based UAV detection system using a LOF algorithm. The RF signals propagated from the UAV flight controller are exploited for the detection of the UAV. However, both Bluetooth and WiFi devices operate at the same frequency band as the UAV controller. Thus, we acquired RF signals from Bluetooth devices, WiFi devices, and UAV controllers. Using WPT, we extracted RF fingerprints (i.e., feature set) from the transient state of the signal. The feature set is used to train a classifier for the novel detection of UAV control signals. In the results, we achieved an average accuracy of $96.7\%$ at 30 dB SNR for the detection of UAV control signals as anomalous signals. More so, it was observed that the performance of the local outlier factor increases with an increase in the number of neighbors. This makes LOF effective at a low SNR. In the future, we will be exploiting other novel detection algorithms and feature extraction methods for more robust performance at a very low SNR.

\bibliographystyle{IEEEtran}
\bibliography{reference}

\end{document}